# Study of epitaxial graphene on non-polar 6H-SiC faces

B.K. Daas[1,a], K.Daniels[1], S.Shetu[1], T.S. Sudarshan[1], and MVS Chandrashekhar[1]

[1]Department of Electrical and Computer Engineering, University of South Carolina

301 S. Main St, Columbia, SC 29208, USA

[a]Daas@email.sc.edu

**Abstract.** We present epitaxial graphene (EG) growth on non-polar a-plane and m-plane 6H-SiC faces where material characterization is compared with that known for EG grown on polar faces. Atomic force microscopy (AFM) surface morphology exhibits nanocrystalline graphite like features for non-polar faces, while the polar silicon face shows step like features. This differing behavior is attributed to the lack of a hexagonal template on the non-polar faces. Non-polar faces also exhibit greater disorder and red shift of all Raman peaks (D, G and 2D) with increasing temperature. This is attributed to decreasing stress with increasing temperature. These variations provide evidence of different EG growth mechanisms on non-polar and polar faces, likely due to differences in surface free energy. We also present differences between a-plane ($11\bar{2}0$) EG and m-plane ($1\bar{1}00$) EG in terms of morphology, thickness and Raman characteristics.

## Introduction

Graphene, a two-dimensional (2D) form of honeycomb crystal structure, is the basic building block of other $sp^2$ carbon nanomaterials, such as nanographite sheets and carbon nanotubes that exhibit unusual electronic and optical properties [1]. Since its discovery, graphene has been produced using mechanical an exfoliation technique whereby graphene layers are peeled off layer by layer using scotch tape [2]. This technique is not suitable for large scale production due to poor yield and repeatability, lack of process control and small sample size. As an alternative, large-area epitaxial graphene (EG) is grown by thermal decomposition of polar c-plane (both Si and C-face) of 4H or 6H SiC in ultra high vacuum or argon environment at high temperature. In this technique, Si sublimes off from the SiC substrate, leaving behind carbon atoms which rearrange themselves into a graphene layer [3]. The growth mechanism of EG on SiC is still a current issue of research because it involves high temperatures with several competing process, such as silicon desorption, carbon diffusion, island nucleation, etc. For the polar c-plane Si face, EG growth is due to the step flow mechanism whereby adjacent steps react with different speeds and the released carbon produces characteristics fingers when a step of higher speed catches up with a slower moving step [4]. Though there is argument about the growth mechanism involved on the C-face [5,6], Tedesco et al. [6] have shown that, argon mediated EG growth on the C-face is defect mediated, as defects are at a higher energy compared to the surface.

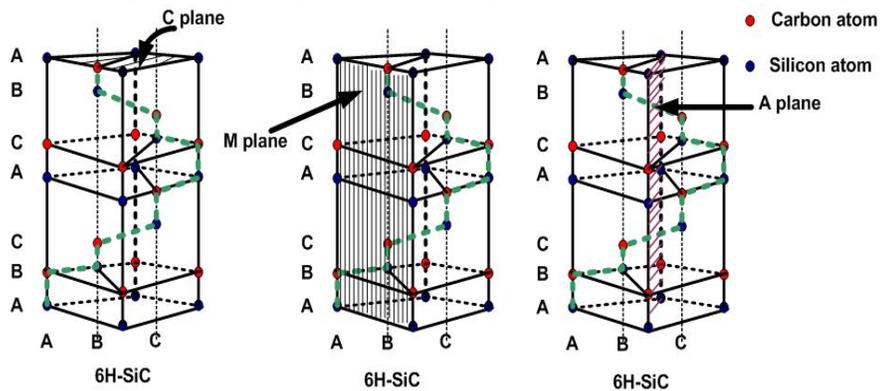

Fig. 1. Crystallographic orientation of 6H-SiC for c, m and a plane, respectively.

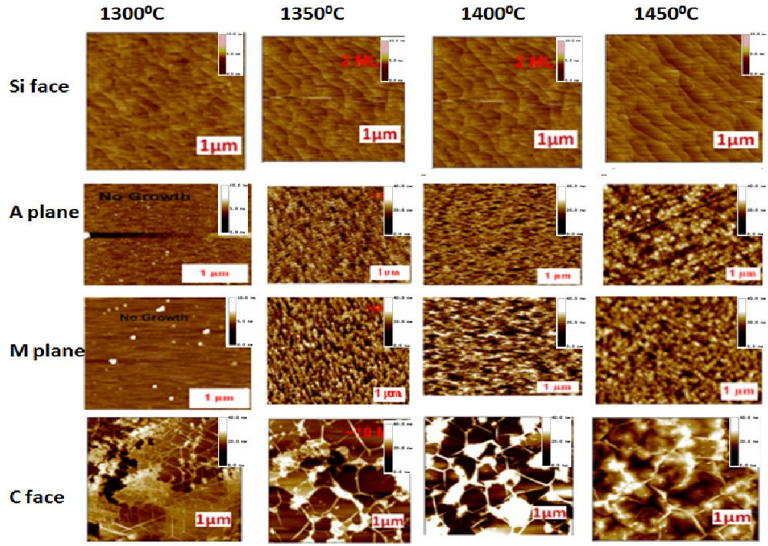

Fig. 2. AFM image of epitaxial graphene grown on a, m and c plane 6H-SiC substrate at three different growth temperatures 1350°C, 1400°C and 1450°C, respectively.

To understand the graphene growth mechanism on SiC and to correlate differences between faces on c-plane SiC, it is essential to study EG growth on a and m plane, EG-a and EG-m (non-polar faces) SiC simultaneously and characterize it in detail. Fig. 1 shows the differences in crystallographic orientation of 6H-SiC a, m and c plane crystal. It is evident that non-polar faces (a and m planes) have more carbon atoms per unit area in a bilayer. These faces also lack a hexagonal template compared to the polar (c plane) Si-face. Difference in crystallographic orientation will be used for explaining graphene morphology and thickness. In this paper we study EG grown on non-polar n-type 6H-SiC, both on a-plane ($11\bar{2}0$), EG-a, and m-plane ($1\bar{1}00$), EG-m, faces, by the well known thermal decomposition technique in an inductively heated RF furnace at high temperature and high vacuum.

**Experimental Details**

Epitaxial growth of large-area graphene by thermal decomposition of commercial <0001> 4H and 6H SiC substrates at high temperature and vacuum [3] can produces EG a few monolayers (ML) to >50 ML thick, depending on growth conditions. In our experiments, EG was grown on a and m plane CMP polished 6H-SiC substrates, nitrogen doped ~$10^{17}$/cm$^3$ with rms roughness <0.75nm measured in a 2.5X2.5μm$^2$ window. The polar c plane Si face was commercially CMP polished with RMS roughness ~0.5nm while the C-face was optically polished with RMS roughness ~1nm. 1cm×1cm$^2$ samples were degreased using Trichloroethylene (TCE), acetone and methanol, respectively. They were then rinsed in DI water for three minutes. The samples were finally dipped in HF for two minutes to remove native oxide and rinsed with DI water before being blown dry with UHP Ar gas. They were then set in the crucible in an inductively heater furnace where high vacuum was maintained (<10$^{-6}$ Torr) and baked out at 1000°C for 13 to 15 hours. We note that no H$_2$ etching was performed on the SiC surface before EG growth. The temperature was slowly raised to the growth temperature (1300-1450°C), using optimized conditions for c plane growth. In every growth, in addition to the a and m plane sample, c plane witness samples were included for comparison. All growths were performed for 60 minutes before cooling to 1000°C at a ramp rate of 7~8°C/min and eventually to room temperature.

Table 1. EG thickness using XPS both for polar and non-polar faces under same growth condition.

| EG grown at 1350°C | Polar face | | Non-polar face | |
|---|---|---|---|---|
| | **Si** | **C** | **m-plane** | **a-plane** |
| No of monolayer | 2 | 9 | 4 | 3 |

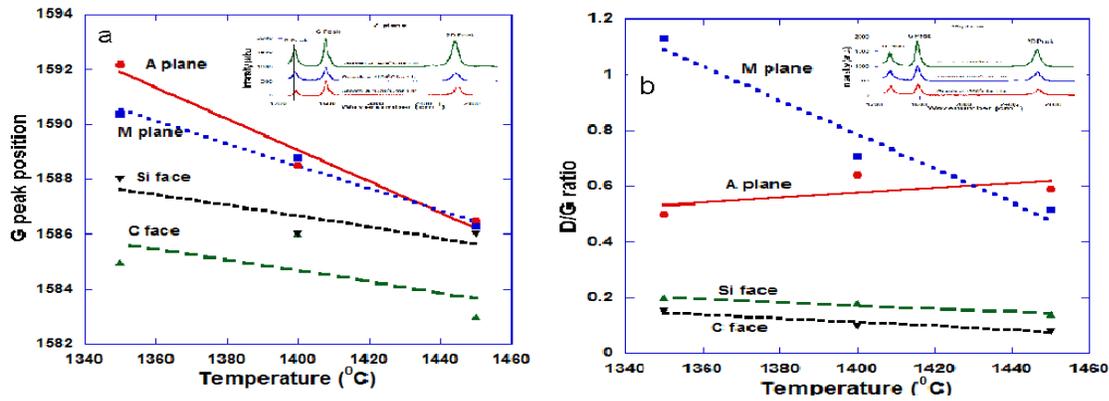

Fig. 3. Variation of G peak and D/G ratio of polar and non-polar faces. Inset: non polar face Raman spectra.

**Results and Discussion**

**Atomic Force Microscopy.** After growth, atomic force microscopy (AFM) measurements in tapping mode were carried out on grown EG-a, EG-m and EG-c to analyze the EG surface and RMS roughness was analyzed within a $2.5 \times 2.5\,\mu m^2$ window for all cases. Fig. 2 shows the AFM of EG-a, EG-m and EG-c grown at temperatures 1300°C to 1450°C. EG-c shows step like features on the silicon face [4] at all growth temperatures (rms roughness ~0.4nm at 1350°C), confirming growth by step flow, that become more prominent with increasing growth temperature. Both EG-a and EG-m exhibit nanocrystalline graphite-like features, which also become more prominent with increasing growth temperature. RMS roughness is ~1nm at 1350°C growth temperature. There appears to be an underlying long-range periodicity that may be from steps on the surface. The poorer morphology of the EG-a and EG-m compared to EG-c is believed to originate from the crystallographic orientation of the substrate, which does not provide a hexagonal template. We restrict our discussion in this paper to a comparison of the material quality and do not comment on the growth mechanism.

**Raman Spectroscopy.** Raman spectroscopy was performed using a micro-Raman setup with laser excitation wavelength at 632nm with a spot size of ~2μm. Raman was used to observe G, D and 2D peaks which are due to in plane-vibrations, disorder and double resonance, respectively [7]. Raman analyses of EG-a and EG-m for different growth conditions are shown as an inset in Fig. 3, whereas with the same growth condition, our previous work shows EG-c with the ratios of intensities of D peak and G peak $I_D/I_G$ <0.2, demonstrating the quality of grapheme on c plane [8]. In Fig. 3, for EG-m and EG-a, the G (1586 to 1590cm$^{-1}$), D (1334 to 1348 cm$^{-1}$) and 2D peaks (2657 to 2678 cm$^{-1}$) confirm the presence of graphene [7,8].

All peaks are red shifted with increasing growth temperature, likely due to thermal stress relief. Fig. 3 shows a shift in the G peak and variation of the $I_D/I_G$ ratio as a function of growth temperature, where corresponding polar face (Si and C face) data are presented for comparison. In Fig. 3(a) the C face shows a lower G peak red shift compared to the Si face, because of thicker layers grown with less stress. EG-a and EG-m exhibit greater red shift in the G peak compared to the polar faces with EG-a showing the greatest shift of all faces. This observation, along with the lower D/G ratio (discussed later), which should increase the G-peak position, indicates greater strain in the EG-a surface (Fig. 3(a)) while the strain is lower in EG-m. The greater G-peak positions for both non-polar faces compared to the polar faces are largely due to the significantly higher disorder in these films [7].

The disorder in the graphitized layers is quantified by the relative peak intensities of Raman D and G peaks, $(I_D/I_G)$ [7], called the disorder ratio. Fig. 3(b) shows a comparison of the disorder ratio as a function of growth temperature for both polar and nonpolar faces. Polar face EG has a lower disorder ratio (<0.2) than nonpolar face EG. For the polar faces, the Si face $I_D/I_G$ is 0.1~0.2, while $I_D/I_G$ <0.05 for the C face. The disorder ratio ($I_D/I_G$) for EG-m decreases from 1.1 to 0.5 as the growth temperature varied from 1350°C to 1450°C, indicating increasing grain size, while $I_D/I_G$ for EG-a does not vary significantly. The reason for this difference is at present unclear, but we speculate that this difference arises from the differences in surface energy [9], leading to different growth mechanisms as evident by the nanocrystalline surface morphology through AFM. Both EG-m and EG-a exhibit much greater disorder than the polar Si-face (<0.1, optimized).

**X-ray Photoelectron Spectroscopy.** X-ray photoelectron spectroscopy (XPS) measurements using a Kratos AXIS Ultra DLD XPS system with a Al K$\alpha$ source. Using XPS we obtained the C1s and Si2p peaks, at normal and 70° beam incidence angles. The graphene C1s peak was normalized to the SiC substrate C1s peak, from which the thickness was determined as described elsewhere [10]. Our result indicates that both EG-m and EG-a layers are thicker than their corresponding polar c-plane Si-face sample while EG-m shows thicker growth than EG-a. These thickness differences (Table 1) can be explained from crystallographic structure shown in fig-1. For c plane, the Si face has one carbon atom at the first bi-layer of thr single crystal, whereas for m and a plane there are five and four respectively (fig-1), to share with the adjacent unit cell. The key here is that we are counting the carbon atom density within a bi-layer, and not only the atoms at the surface. This number is the same on the C-and Si faces, as the arrangement on those faces is the same. The arrangement on the a/m planes is different, leading to a different density of carbon atoms/bi-layer. This indicates higher carbon atom density in a bi-layer of a and m plane compared to Si face. Higher carbon atom density at the surface provides possibility of more carbon atoms to rearrange themselves as Si sublimes off during EG. This is reflected through the thickness analysis (shown in Table 1), where the c plane Si face has 2 ML whereas the a plane and m plane has 3 ML and 4 ML, respectively at 1350°C growth temperature. This thickness difference may also be due to the defective nature of the non-polar EG layers, allowing more Si escape from the surface.

**Conclusion**
In summary, we have demonstrated EG growth on polar c-plane silicon and carbon faces while non-polar face EG was compared with the polar faces in surface morphology and graphitization quality. We believe that nanocrystalline graphite features on non-polar faces originate from the lack of a hexagonal template in the substrate. Raman measurements shows polar faces exhibit better graphitization quality than non-polar faces. Non-polar face graphitization is thicker than polar Si face as confirmed through XPS. Comparing the non-polar face graphitization, between a and m planes, it is evident that a-plane graphitization is better quality than m plane graphitization, as determined by Raman, while m-plane graphitization is thicker than a-plane. In this paper, we focus on comparing graphitization quality between polar and non-polar faces, rather than the growth mechanism. Further investigation is underway to clarify the growth mechanism on non-polar faces.

**Acknowledgement**
The authors would like to acknowledge the Southeastern Center for Electrical Engineering Education, and the National Science Foundation (NSF) ECCS-EPDT Grant #1029346 under the supervision of program director Rajinder Khosla for funding this work.